Original Article

# Processing HSV Colored Medical Images and Adapting Color Thresholds for Computational Image Analysis: a Practical Introduction to an open-source tool

Lie Cai, MD[1]; André Pfob, MD[1,2*];

[1] Department of Obstetrics and Gynecology, Heidelberg University Hospital, Heidelberg, Germany

[2] National Center for Tumor Diseases (NCT) and German Cancer Research Center (DKFZ), Heidelberg, Germany

* corresponding author

André Pfob, MD

Department of Obstetrics & Gynecology, Heidelberg University Hospital

Im Neuenheimer Feld 440

69120 Heidelberg, Germany

+49 6221 56 310290

Andre.pfob@med.uni-heidelberg.de

Twitter handle

@andrepfob

@liecaii

**Abstract**

**Background**

Using artificial intelligence (AI) techniques for computational medical image analysis has shown promising results. However, colored images are often not readily available for AI analysis because of different coloring thresholds used across centers and physicians as well as the removal of clinical annotations. We aimed to develop an open-source tool that can adapt different color thresholds of HSV-colored medical images and remove annotations with a simple click.

**Materials and Methods**

We built a function using MATLAB and used multi-center international shear wave elastography data (NCT 02638935) to test the function. We provide step-by-step instructions with accompanying code lines.

**Results**

We demonstrate that the newly developed pre-processing function successfully removed letters and adapted different color thresholds of HSV-colored medical images.

**Conclusion**

We developed an open-source tool for removing letters and adapting different color thresholds in HSV-colored medical images. We hope this contributes to advancing medical image processing for developing robust computational imaging algorithms using diverse multi-center big data. The open-source Matlab tool is available at [https://github.com/cailiemed/image-threshold-adapting](https://github.com/cailiemed/image-threshold-adapting).

**Highlights**

An open-source tool that can adapt different color thresholds of HSV-colored medical images.

The newly developed pre-processing Matlab function successfully works on multi-center, international shear wave elastography data (NCT 02638935).

Step-by-step instructions with accompanying code lines were provided, easy to follow and reproduce.

Abbreviations

AI Artificial Intelligence

HSV Hue Saturation Value

RGB Red Green Blue

SWE Shear Wave Elastography

VTIQ Virtual Touch Tissue Imaging

KNN K-nearest neighbors

# 1. Introduction

Recently, integrating artificial intelligence (AI) techniques into medical imaging has shown impressive results and may have great potential to transform the diagnostic process.(Ricci Lara et al., 2022) Colored medical images (e.g. shear wave elastography images, cardiovascular images, thermodynamics images) contain more information than simple gray-level images. Numerous colored medical images are produced daily during clinical routine, and retrospectively mining data from them can advance the development of imaging algorithms.

Many researchers have tried to develop various algorithms using colored medical images to answer clinical questions,(Chen et al., 2022) however, in practice, the number of colored images readily available for computational image processing is very limited due to the following reasons: (1) The color/intensity threshold can be individually adjusted, resulting in a range of images where the same color is associated with a different range of values (e.g. if the maximum threshold of an Elastography image can range from 1 to 10m/s with color-coded from blue to red – if the threshold is set to 6m/s, the color range is adapted from blue to red on the new 1 to 6m/s scale). The different thresholds make such color images not readily available for computational image analyses. (2) The majority of medical images contain annotations and letters made by clinicians for clinical interpretation. Previous studies have developed algorithms to remove letters on gray-level images based on color threshold identification. (Tustison et al., 2010; Wang et al., 2022) However, due to the different color distribution, these algorithms are not readily available for colored medical images.

The color distribution of medical images often matches the Hue-Saturation-Value (HSV) distribution, first described by Alvy R Smith(Smith, 1978). The HSV color distribution is different from the commonly known Red-Green-Blue (RGB) distribution (see Figure 1). This study aims to develop an open-source tool that can adapt different colored images' thresholds and remove annotations and letters with a simple click from HSV-colored medical images. With this, we want to contribute to advancing medical image processing for developing robust algorithms using diverse multi-center big data.

## 2. Materials and methods

*2.1 Dataset*

We used multi-center international shear wave elastography (SWE) data (NCT 02638935).(Pfob et al., 2022) This multi-center trial aimed to evaluate the diagnostic performance of SWE for breast cancer diagnosis. Data was documented from 12 centers in 7 countries. These images showed a great variance of the maximum shear wave velocity setting made by examiners, ranging from 0.5m/s to 10m/s. In this study, we selected an image with a shear wave velocity of 10m/s and nearly all colors range from 0.5m/s to 10m/s as the reference image. (Figure 1)

Figure 1 Reference image and its color distribution in RGB format and HSV format.

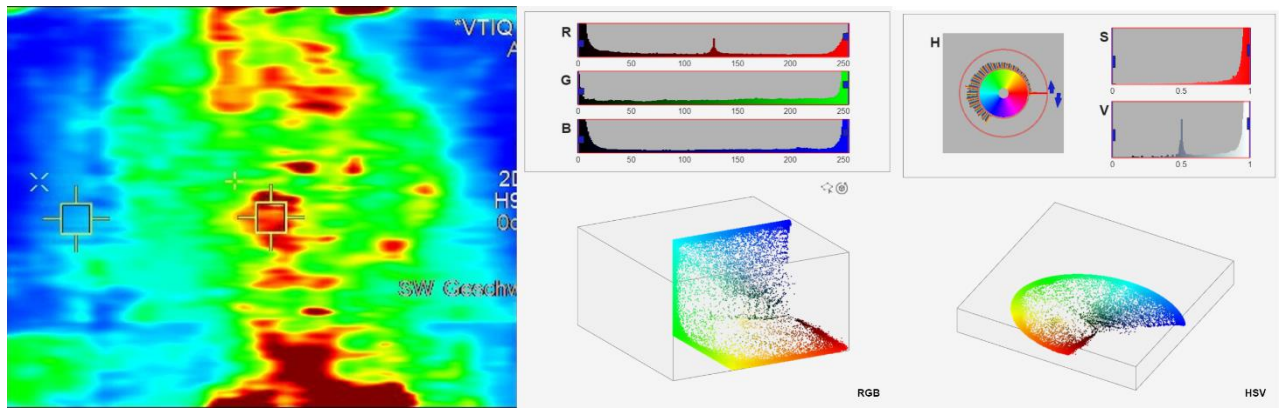

The letters shown in the Figure are parameters of the shear wave elastography machine and do not reveal any information about patients.

2.2 Software

MATLAB R2023b was used for all analyses.

2.3 Image processing

The whole image processing steps were integrated into one function named "pre_process_HSV" in MATLAB. Readers who want to reproduce the results or use this function to process their own images can simply download them from https://github.com/cailiemed/image-threshold-adapting and call it. The function has two output images: an image with removed letters, and the color-scale adapted image. For better clarification, we will introduce the inner codes of the function step by step.

2.4 Fill dark spaces with blue color

Some images of the VTIQ dataset contain artifacts, showing as purely black spaces. While black does not exist on colored SWE images, putting these dark artifacts into subsequent analysis may produce bias. We replaced these dark spaces with blue color, which stands for low shear wave velocity (Table 1, task 1). We selected the dark threshold as 0.148 for HSV's V channel using Matlabs "ColorThresholder" (Table 1, task 1.3). Then we created a dark mask with V values lower than the dark threshold. (Table 1, task 1.4) We replaced the dark spaces from the masked image with light green-blue colors [0.606, 1.000, 1.000]. Figure 2 shows the original image with artificial dark spaces and the corrected image.

Figure 2 The original image with dark spaces and the corrected image.

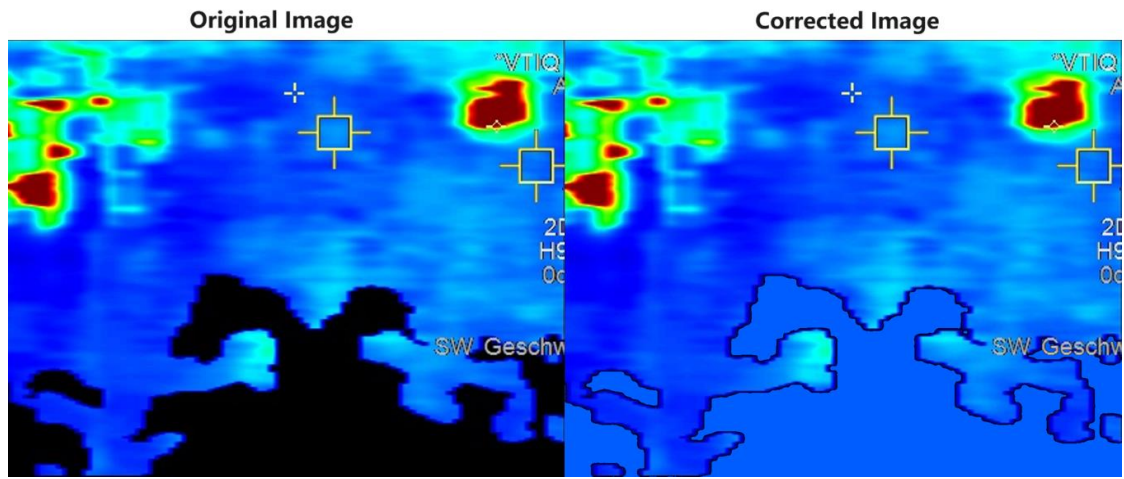

The letters shown in the Figure are parameters of the shear wave elastography machine and do not reveal any information about patients.

*2.5 Remove letters*

Using the ColorThresholder to adjust the HSV's S value, we selected the best minimum value equal to 0.700 to exclude letters (Table 1, task 2.1). Then we created a mask based on the specified channels' threshold (Table 1, task

2.2). We initialized the masked image based on the input image, set the background pixels where the mask is false to zero, and then inverted the mask to select the letters. (Table 1, tasks 2.3 and 2.4) However, if we strictly remove letters and replace them with colors, it will cause clear edges that still show the outline of the letters. To expand the mask, we used the dilate morphology operation from the App ImageSegmenter, with a disk shape and a radius of 6 (Table 1, task 2.6). Then we replaced the expanded mask's color with black in the masked image (Table 1, task 2.7). Next, we split the expanded masked image into red, green, and blue channels in RGB format. We identified the indexes of rows and columns whose values in the RGB three channel are all equal to 0, and named them "missingpositions" (Table 1, task 2.8) Here we presented an exemplary image to show why we need to find the indexes of values from three channels that are all equal to 0. In Figure 3 there's a black capital letter "E" in the center, segmenting it in a separate channel does not work, only a combined RGB channel can successfully segment it.

Figure 3 Exemplary image of segmenting letters in different RGB channels.

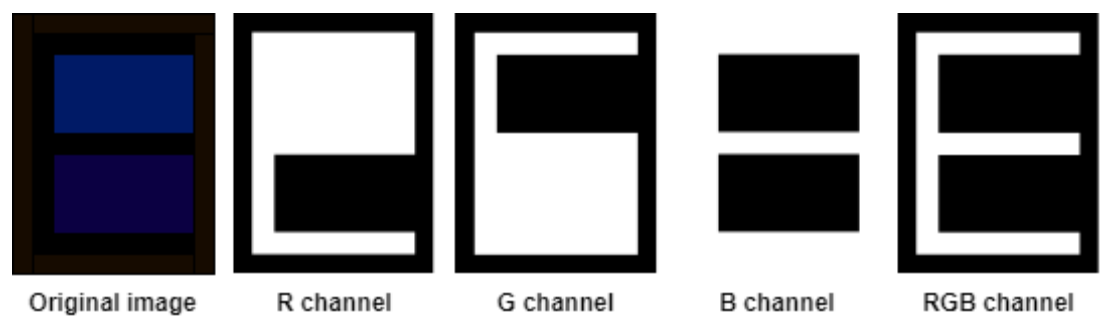

Then we replaced the black masks with missing values (NaN) in all three channels separately (Table 1, task 2.9) and calculated all the missing values by using the K-nearest neighbors (KNN) algorithms in separate channels (Table 1, task 2.10). As the optimal k value may depend on the specific use-case, we inserted a numeric slider control range from 5 to 40, allowing to adjust the value of k. We rebuilt the image of the removed letters named "imgnoword" (Table 1, task 2.11). Figure 4 shows an example of removing letters.

Figure 4. The original image and the image of removed letters.

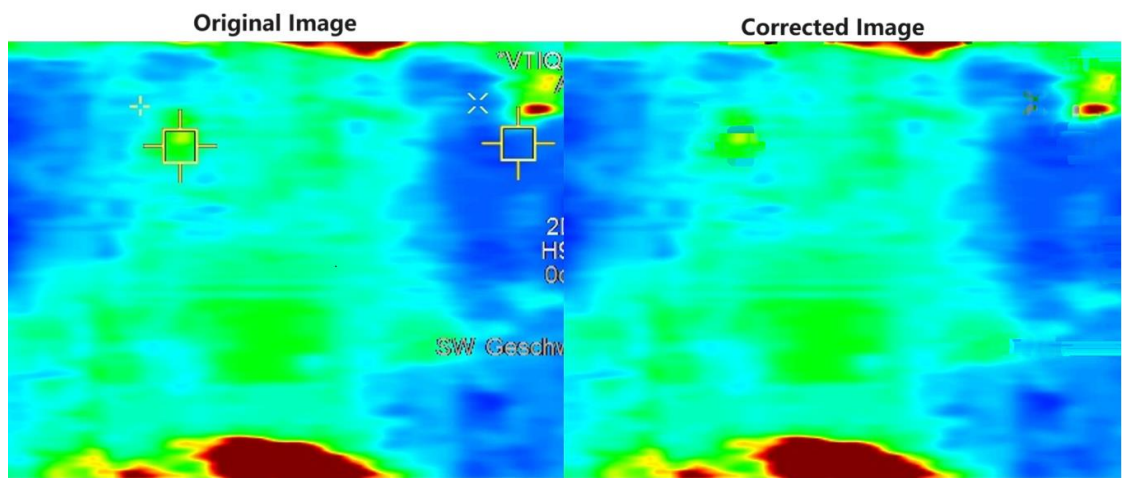

The letters shown in the Figure are parameters of the shear wave elastography machine and do not reveal any information about patients.
The "Fill dark spaces" step is also implemented in this image, but our algorithm has no negative effects on pure-colored images.

*2.6 Threshold adapting*

First, we converted the "imgnoword" from RGB to HSV, then split it into H, S, and V channels (Table 1, task 3.1). Checking the reference color bar, we see its color begins from blue to red, which matches the distribution of HSV. Table 2 presents the color distribution of the reference color bar and corresponding H values. After converting the color reference bar from RGB to HSV (Figure 5), we found on the top of the bar some yellow plaques called noises, with H values ranging around 0.99, it's because some dark red colors were near chestnut color, whose H value was near 1. Except for noises, the whole color bar's H value ranges from blue (0.667) to red (0), we removed noises by converting them to red (0.001) color. (Table 1, task 3.2)

Figure 5 Noises in the color reference bar after the transition.

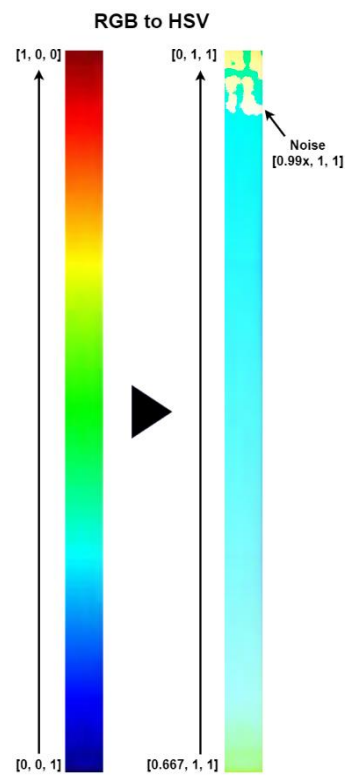

After converting the "imgnoword" from RGB to HSV, it has a linear distribution of color based on the H value, we set the shear wave velocities value as the Y axis, with a minimum value of 0.5m/s, and the colors' H values as the X axis, ranging from 0 to 0.667. Then the curve of shear wave velocities changes with the H values of all images (including the reference image) and

can be expressed by the following linear function:

$$Y = aX + b$$

For the reference images with a max shear wave velocity of 10m/s, its curve passes (0, 10) and (0.667, 0.5), after calculation, its function is as follows:

$$Y_1 = -14.25X_1 + 10$$

Here we present an example of a test image with a max shear wave velocity of 6.5m/s, its curve passes (0, 6.5) and (0.667, 0.5), and its function is as follows:

$$Y_2 = -9X_2 + 6.5$$

When adapting the test image's color to the reference image, we just need to calculate the $X_1$ value when the two curves have the same Y value, calculation is as follows:

$$Y_1 = Y_2$$

$$-14.25X_1 + 10 = -9X_2 + 6.5$$

$$X_1 = (3.5 + 9X_2)/14.25$$

Figure 6a puts the two curves in one $X$-$Y$ coordinate, Figure 6b shows the image adapting of the two curves.

Figure 6 the two curves in the $X$-$Y$ coordinate

Figure 6a diagram of the reference image curve and the test image curve.

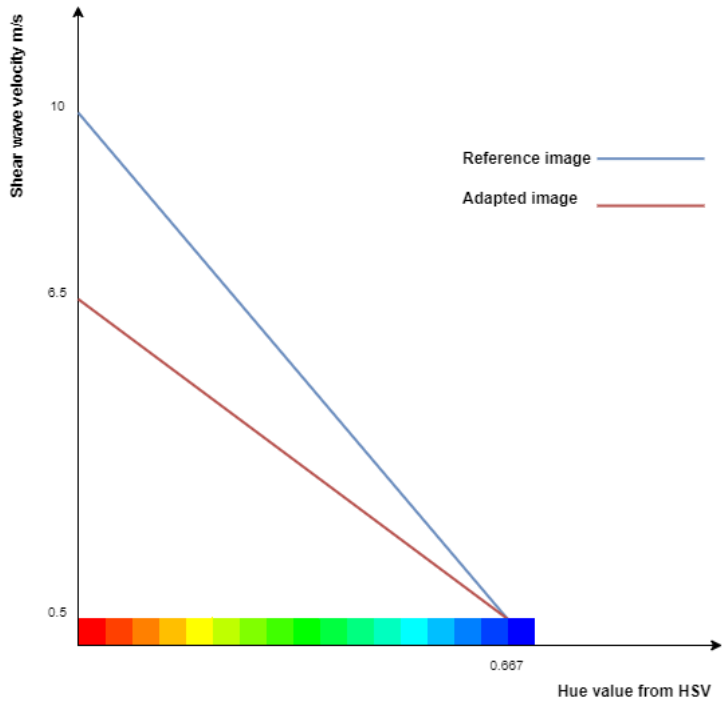

Figure 6b The image adapting of the two curves.

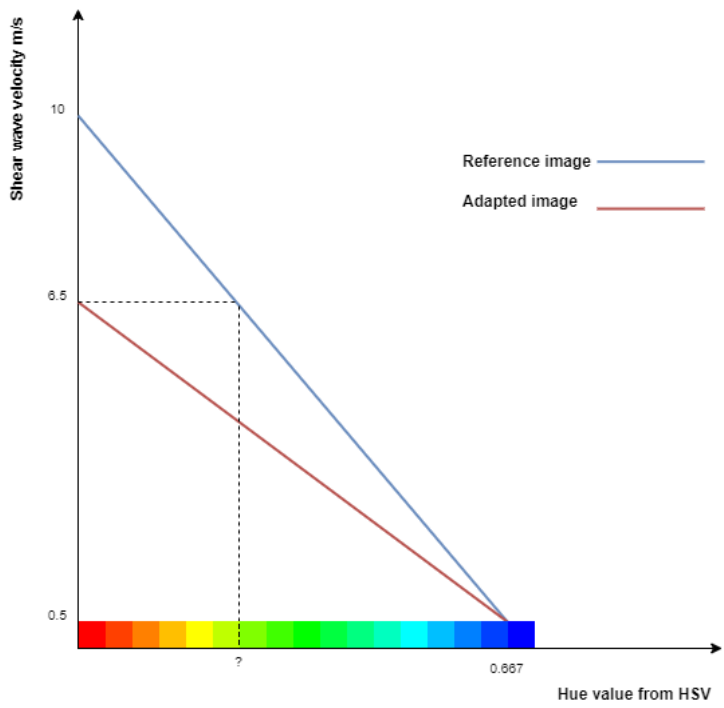

For all images, we set the max shear wave velocity from the reference image as *m*, the max shear wave velocity from the test image as *t*, the original H value as *h*, and the adapted H value as *h'*, the calculation can be expressed as follows (Table 1, task 3.3):

$$h' = ((1.5*t - 0.75)*h - t + m)/((m-0.5)*1.5)$$

Figure 7 shows the example of color adapting from the test image (max: 6.5m/s) to the reference image (max: 10m/s).

Figure 7 Color adapting.

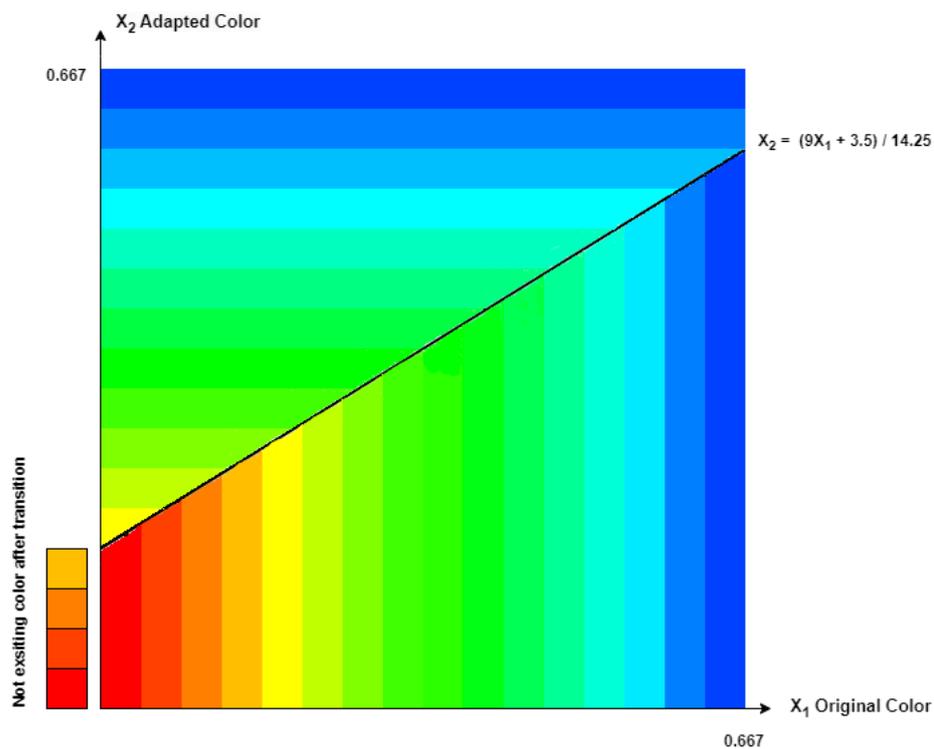

The curve splits the square into two parts, the lower half part represents the whole color distribution of the test image, and the upper half part represents the adapted color distribution, after transition, some colors do not exist and are presented in a separate column in the lower left corner.

After customizing the test image's H value to the reference image, we set the test image's S and V values that were originally not equal to 1 equal to 1 (Table 1, task 3.4). This aimed to make all adapted colors in the standard form, Figure 8 shows an example without color standardization, it shows a dark green color in stiff tissues, which does not exist in the color bar.

Figure 8 Color adapting without standardization.

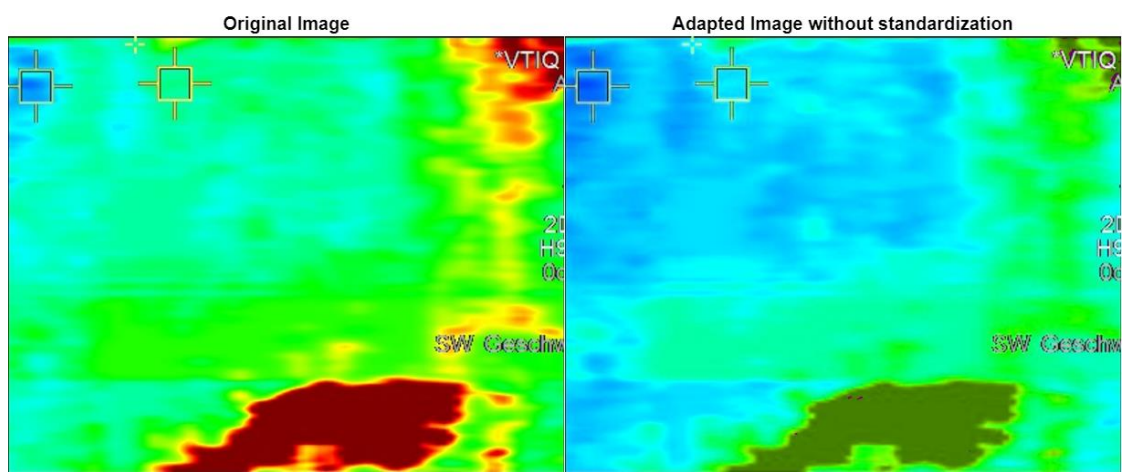

The letters shown in the Figure are parameters of the shear wave elastography machine and do not reveal any information about patients.

Then we rebuilt the image by combining the H, S, and V channels, and the image adapting was finished.

## 3. Results

We demonstrate exemplary images with different max shear wave velocities from the VTIQ dataset that were pre-processed with our newly developed function in Figures 9 and 10. We demonstrate that the pre-process function successfully removed letters and adapted to the reference image with 10m/s of the max shear wave velocity. Figure 9 shows the original images, the images with removed letters, and the threshold-adapted images. Figure 10 shows the results of images with artifacts.

Figure 9 Exemplary images of removing letters and adapting the color threshold.

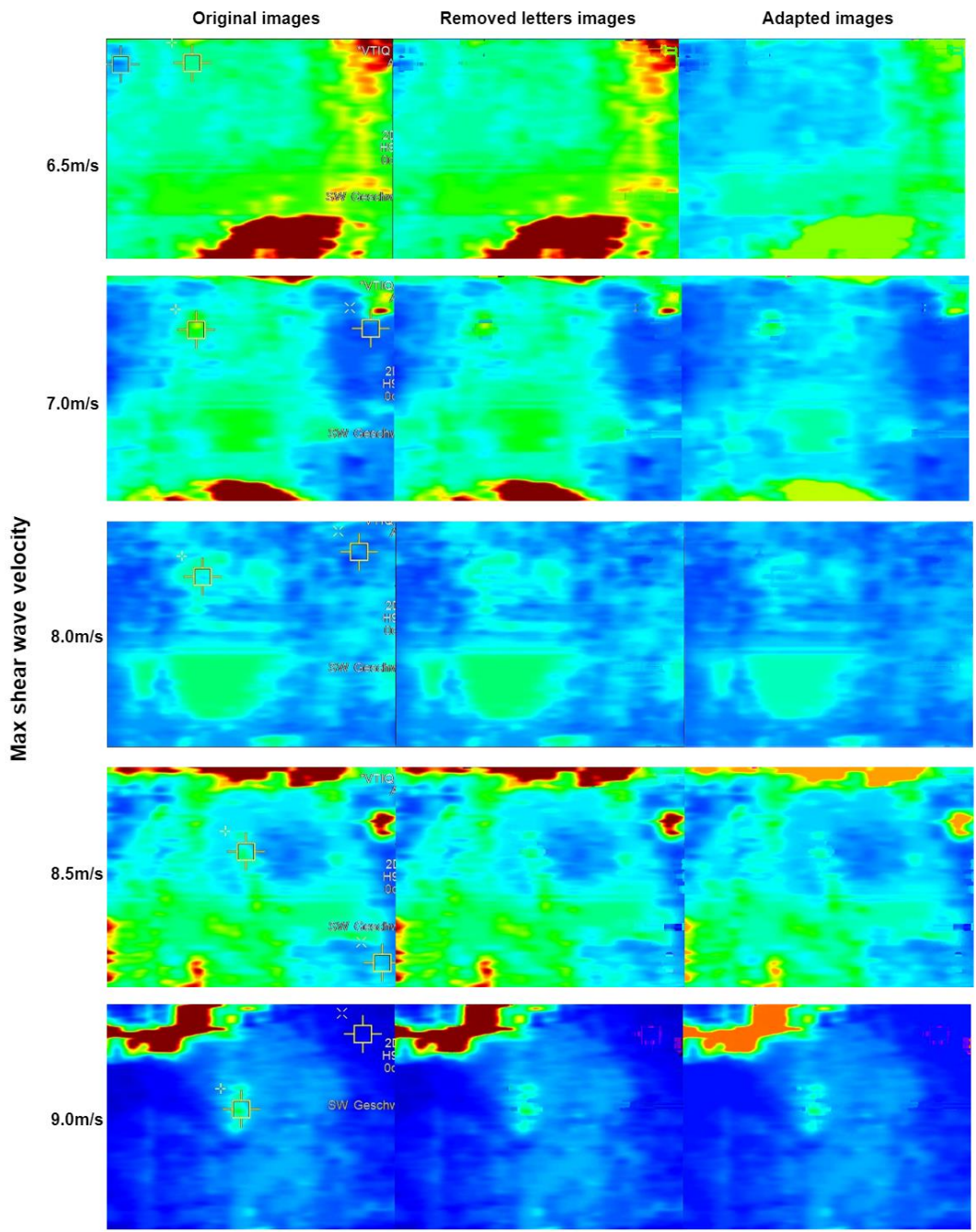

The letters shown in the Figure are parameters of the shear wave elastography machine and do not reveal any information about patients.

Figure 10 Exemplary images with artifacts of removing letters and adapting the color threshold.

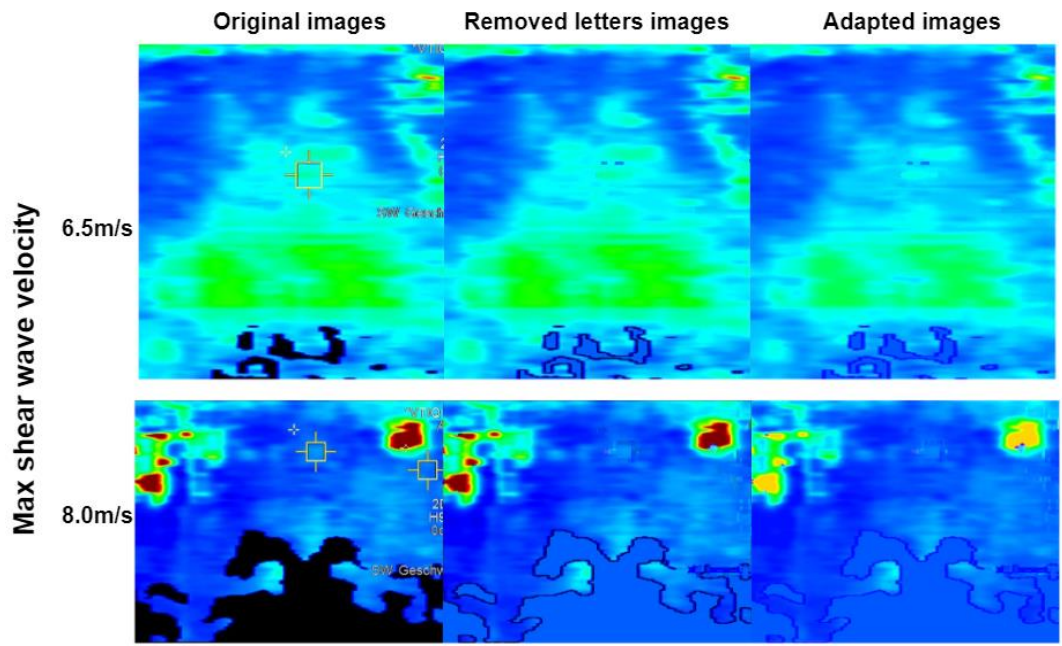

The letters shown in the Figure are parameters of the shear wave elastography machine and do not reveal any information about patients.

## 4. Discussion

In this analysis, we developed an open-source tool for removing letters and adapting different color thresholds in HSV-colored medical images. To the best of our knowledge, this is the first analysis trying to solve the pre-annotations and diverse color thresholds in multi-center, HSV-colored medical imaging datasets. The function produced stable output images, even considering artifacts in the original images.

Using colored medical images for AI-based analysis is a pressing issue. There are several applications where the interpretation of color-coded images is very challenging to the human eye, leading to inter-operator variability and limited diagnostic performance. For example, shear-wave Elastography has been evaluated for breast cancer diagnostics for more than 20 years but has been limited so far by issues regarding inter-operator variability and image interpretation.(Golatta et al., 2023, 2022) However, large-scale AI-based analysis of shear-wave images has been limited so far by variations in the image acquisition process, resulting in shear-wave images that have different color thresholds. We hope that this open-source tool contributes to advancing HSV medical image processing for developing robust computational imaging algorithms using diverse multi-center big data.

## 5. Conclusion

We developed an open source tool for removing letters and adapting different color thresholds in HSV colored medical images. We hope this contributes to advancing medical image processing for developing robust computational imaging algorithms using diverse multi-center big data. The open-source Matlab tool is available at [https://github.com/cailiemed/image-threshold-adapting](https://github.com/cailiemed/image-threshold-adapting).

**Acknowledgement**

None.

**Funding**

This research did not receive any specific grant from funding agencies in the public, commercial, or not-for-profit sectors.

Table 1 Summary of key Tasks and accompanying MATLAB codes for Imaging Processing.

| Task | Code in MATLAB |
| --- | --- |
| 1) Fill dark spaces with blue color | |
| 1.1) Import a test image called "X" and convert it from uint8 to double | ```X = imread("test image.png")```<br>```X = im2double(X);``` |
| 1.2) Convert X from RGB to HSV | ```X = rgb2hsv(X);``` |
| 1.3) Set the threshold of dark spaces by Color Thresholder | ```darkThreshold = 0.148;``` |
| 1.4) Specifying the dark mask in HSV's V channel | ```vChannel = X(:,:,3);```<br>```darkMask = vChannel < darkThreshold;``` |
| 1.5) Replace dark spaces with specifying blue color | ```blueColor = [0.606, 1.000, 1.000];```<br>```for channel = 1:3```<br>`    channelData = X(:,:,channel);`<br>`    channelData(darkMask) = blueColor(channel);`<br>`    X(:,:,channel) = channelData;`<br>```end``` |
| 2) Remove letters | |
| 2.1) Specifying the threshold value of X by Color Thresholder | ```channel1Min = 0.000;```<br>```channel1Max = 1.000;```<br>```channel2Min = 0.700;```<br>```channel2Max = 1.000;``` |
| | `channel3Min = 0.000;`<br>`channel3Max = 1.000;` |
| 2.2) Create a mask based on the specified channels' threshold | `sliderBW = (I(:,:,1) >= channel1Min ) & (I(:,:,1) <= channel1Max) & ...`<br>`    (I(:,:,2) >= channel2Min ) & (I(:,:,2) <= channel2Max) & ...`<br>`    (I(:,:,3) >= channel3Min ) & (I(:,:,3) <= channel3Max);`<br><br>`BW = sliderBW;` |
| 2.3) Initialize the masked image based on the input image, and set the background pixels where BW is false to zero. | `maskedRGBImage = X;`<br><br>`maskedRGBImage(repmat(~BW,[1 1 3])) = 0;` |
| 2.4) Invert the mask | `BW = im2double(BW);`<br>`BW = imcomplement(BW);` |
| 2.5) Set the parameters of the morphology operation by Image Segmenter | `radius = 6;`<br>`decomposition = 0;`<br><br>`se = strel('disk', radius, decomposition);` |
| 2.6) Expand the mask | `expandedMask = imdilate(BW, se);`<br>`expandedMask = logical(expandedMask);` |
| 2.7) Set the color of the tested image where in the expanded mask to black | `X(repmat(expandedMask, [1, 1, 3])) = 0;` |
| 2.8) Split the test image to RGB matrix, find the indexes of rows and columns that values in the RGB three channel all equal to 0, and name them as missingpositions. | `[R, G, B] = imsplit(X);`<br><br>`[zrows, zcols] = find(R == 0 & G == 0 & B == 0);`<br>`R = im2double(R);`<br>`G = im2double(G);`<br>`B = im2double(B);`<br><br>`missingPositions = [zrows, zcols];` |

| 2.9) Replace missingpositions to NaN in RGB three channels separately | ```matlab
for i = 1:size(missingPositions, 1)
    row_coord = round(missingPositions(i, 1));
    col_coord = round(missingPositions(i, 2));

    R(row_coord, col_coord) = NaN;
end

for i = 1:size(missingPositions, 1)
    row_coord = round(missingPositions(i, 1));
    col_coord = round(missingPositions(i, 2));

    G(row_coord, col_coord) = NaN;
end

for i = 1:size(missingPositions, 1)
    row_coord = round(missingPositions(i, 1));
    col_coord = round(missingPositions(i, 2));

    B(row_coord, col_coord) = NaN;
end
``` |
|---|---|
| 2.10) Compute the missingpositions separately in RGB three channels using the K-nearest neighbor algorithm | ```matlab
rImputed = knnimpute(R, k);
gImputed = knnimpute(G, k);
bImputed = knnimpute(B, k);
``` |
| 2.11) Rebuild the image of removed letters as imgnoword | ```matlab
imgnoword = cat(3,rImputed, gImputed, bImputed);
``` |

3) Threshold adapting

3.1) Convert imgnoword from RGB to HSV, and split it into three channels

```
imgnoword = rgb2hsv(imgnoword);
[h,s,v] = imsplit(imgnoword);
```

3.2) Remove noise that is near the max threshold

```
h(h>0.80) =0.001;
```

3.3) Calculate the adapted H value

```
h = ((1.5*t - 0.75)*h - t + m)/((m-0.5)*1.5);
```

3.4) Convert S and V channels to the standardized value

```
v(v~=1)=1;
s(s~=1)=1;
```

3.5) Rebuild the adapted image and convert it back to RGB

```
imgadapt = cat(3,h,s,v);
imgadapt = hsv2rgb(imgadapt);
```

Table 2. The color distribution of the reference color bar and corresponding H values.

| Color Name | Color | H degrees | H values |
|---|---|---|---|
| Red | | 0 | 0.000 |
| Orange red | | 15 | 0.042 |
| Orange | | 30 | 0.083 |
| Orange yellow | | 45 | 0.125 |
| Yellow | | 60 | 0.167 |
| Green yellow | | 75 | 0.208 |
| Grass green | | 90 | 0.250 |
| Yellow green | | 105 | 0.292 |
| Bright green | | 120 | 0.334 |
| Light green | | 135 | 0.375 |
| Spring green | | 150 | 0.417 |
| Green cyan | | 165 | 0.458 |
| Cyan | | 180 | 0.500 |
| Blue cyan | | 195 | 0.542 |
| Green blue | | 210 | 0.569 |
| Light blue | | 225 | 0.611 |
| Blue | | 240 | 0.667 |

**Data sharing statement**

*Will individual participant data be available (including data dictionaries)?*

Yes

*What data in particular will be shared?*

We will provide the reference image and all the test images free for readers to reproduce the results.

*What other documents will be available?*

none

*When will data be available (start and end dates)?*

Immediately following publication. No end date.

*With whom?*

All readers.

*By what mechanism will data be made available?*

To gain full access to the VTIQ dataset, proposals should be directed to [andre.pfob@med.uni-heidelberg.de](mailto:andre.pfob@med.uni-heidelberg.de), data requestors will need to sign a data access agreement.

**Code availability**

We used the MATLAB R2023b programming language. The function and the implementation codes are available at https://github.com/cailiemed/image-threshold-adapting.